\documentclass[journal=jpcafh,manuscript=article]{achemso}
\usepackage{amsmath}
\usepackage{amssymb}
\usepackage{graphicx}
\usepackage{color}
\usepackage{multirow}
\usepackage{natbib}

\newcommand{\be}{\begin{equation}}
\newcommand{\ee}{\end{equation}}
\newcommand{\bea}{\begin{eqnarray}}
\newcommand{\eea}{\end{eqnarray}}

\def\rrscan{\ensuremath{{\mathrm{r}{^2}\mathrm{SCAN}}}}

\newcommand{\ofr}{\ensuremath{(\mathbf{r})}}

\newcommand{\half}{{1/2}}

\title{Spin-Crossover From a Well-Behaved, Low-Cost meta-GGA Density Functional}
\author{Daniel Mej{\'i}a-Rodr{\'i}guez}
\email{dmejiarodriguez@ufl.edu}
\author{S.B.\ Trickey}
\email{trickey@qtp.ufl.edu}
\affiliation{Center for Molecular Magnetic Quantum Materials,
Quantum Theory Project, Department of Physics, University of Florida,
Gainesville, FL 32611\\
29 Sept. 2020}

\begin{document}

\begin{abstract}
  \noindent The recent major modification, $\rrscan$,  of the SCAN
  (strongly constrained and
  appropriately normed) meta-GGA exchange-correlation functional
  is shown to give substantially better spin-crossover electronic energies
  (high spin minus low spin) on a benchmark data set than the original
  SCAN as well as on some Fe complexes. The deorbitalized counterpart $\rrscan$-L is almost as good as SCAN and
  much faster in periodically bounded systems. A combination strategy for 
  balanced treatment of molecular and periodic spin-crossover therefore is
  recommended.
\end{abstract}

\section{Introduction}
\textit{Context} - The essential physical trait of a spin-crossover (SCO)
molecule is a small energy difference between the ground state of one
spin and an excited state of a different spin.  Small in this context
typically means a few kcal/mol (i.e. a few hundred meV).  Calculation
of such differences is challenging.  An added challenge is that 
spin-crossover is of greatest interest in condensed phases.  Predictive
calculation protocols therefore must be equally accurate for both isolated
molecules and their condensed phases.

Meeting that challenge has been difficult.  It is not our purpose to
survey the literature.  For that, see
Refs.\ \citenum{IoannidisKulik2015,MortensenKepp2015,CireraRuiz2016,HardingHardingPhonsri2016,Kepp2016,AmalbinoDeeth2017,Flores-LeonarEtAl2017,SirirakEtAl2017,CireraNadalRuiz2018}.
The last-mentioned of these is particularly relevant.  It presented
a data-base of 20 molecules in which SCO arises from a first-row
transition metal.  Against that data-base, the authors of
Ref. \citenum{CireraNadalRuiz2018} tested several rather sophisticated
density functional approximations (DFAs) for exchange and correlation
(XC) and concluded that the hybrid Tao-Perdew-Staroverov-Scuseria (TPSSh) 
\cite{TPSSh,TPSS} DFA was best overall.

The focus on DFAs stems from the need for affordable calculations both
on large molecules and on their condensed aggregates.  Refined
wave-function methods are applicable, though costly, in SCO molecules.
They are prohibitively costly in the condensed phases.  In principle,
density functional theory (DFT) 
methods should be applicable to both.  Until recently, however, all
affordable, ``lower-rung'' \cite{PerdewSchmidt2001} DFAs have exhibited
bias either to the molecular or the condensed side.

The recommendation of TPSSh is itself somewhat problematic.  The drawback
that is relevant here is its hybrid character, namely, inclusion of
10\% single-determinant exchange (often inaccurately called Hartree-Fock
or exact exchange; both terms have precise, well-defined meanings that
are not met by a hybrid DFA). 

The non-hybrid antecedent of TPSSh, TPSS,
is a meta-Generalized Gradient Approximation (meta-GGA).  In
meta-GGAs, chemically distinct  electron density inhomogeneities
are recognized by use of so-called indicator functions. In the
case of TPSS there are two.  Based on their 
values, the meta-GGA switches between a non-empirical GGA DFA that is
constructed to work well with molecular-like environments
and another for condensed-phase environments.

Largely for reasons of accuracy, TPSS has been supplanted by a more
refined meta-GGA called SCAN, for ``strongly constrained and
appropriately normed'' \cite{SCAN,SCANNature}.  It uses only one
indicator function, denoted $\alpha\ofr$.  With comparatively few
exceptions (e.g.  Ref. \citenum{DMR-SBT-SCANmag}) SCAN has proven
successful in predicting a wide variety of molecular and condensed
phase properties.  That success is a consequence of the physical
realism associated with enforcement in SCAN of all the rigorous
constraints that a meta-GGA can meet, along with calibration to the
energies of selected primitive physical systems (the ``appropriate
norms''; see Supplemental Material to Ref. \citenum{SCAN}).

\textit{SCAN and spin crossover} - Motivated by other successful uses 
of SCAN, Cirera and Ruiz \cite{CireraRuiz2020} tested it recently against  
the 20-molecule SCO data-base of Ref. \cite{CireraNadalRuiz2018}.  Their conclusion was that SCAN ``\ldots gives the right
ground state for the whole set of test cases'' and is ``\ldots the
unique pure DFT functional to provide with comparable results for such
a challenging test.''  All of the systems have low-spin (LS) as
the ground state, as indeed is found by SCAN.  However, the SCO energy
differences
\be
\Delta E_{HL} := E_{H}-E_{L}  \; ,
\label{deltaEHSLS}
\ee
with $E_H$ ($E_L$) the high-spin (low-spin) total energy from
SCAN were only semi-quantitative at best. In some cases they are off
by as much as a factor of 2 or more.  Note that these comparisons were
with respect to TPSSh results for $\Delta E_{HL}$. Those values themselves  
lead to an overestimation of the crossover temperature
\cite{CireraNadalRuiz2018}. 
A technical difficulty
is that SCAN calculations required dense radial integration grids.
An uncomfortable aspect is that the best range of $\Delta E_{HL}$ values from
SCAN were generated with a sub-optimal (not fully converged) grid.  

Grid density and SCF convergence difficulties with SCAN already had  become
well-known among practitioners.  Those problems were addressed by  Bart\'ok
and Yates \cite{rSCAN} with regularized SCAN (rSCAN).  It refined $\alpha$
and smoothed the SCAN switching function to yield improved computational
behavior.   Though rSCAN  preserves 
the good molecular bond lengths and vibrational frequencies given by 
SCAN, it sacrifices SCAN performance for benchmark molecular heats of formation
\cite{rSCANcomment}.  In periodic solids, SCAN and rSCAN are about the
same for lattice constants and cohesive energies 
\cite{rSCANcomment} on a 55 solid test set \cite{SCANrVV10}
and for bulk moduli on a 44 solid set \cite{TranStelzlBlaha}.

Very recently Furness et al. \cite{FurnessEtAl2020} have cured the
deficiencies of rSCAN by constructing a similar regularization that
restores all but one of the constraints satisfied by SCAN but violated by rSCAN.
The regularized-restored SCAN functional ($\rrscan$) that results
combines the strong performance trends of SCAN relative to molecular
and solid data sets with the numerical stability of rSCAN.

A separate conceptual and computational issue of meta-GGAs in general
is their explicit dependence upon the Kohn-Sham (KS) orbitals.
As a matter of
practice, the computational costs from that dependence leads to the use 
of the generalized KS (gKS) equations rather than the multiplicative
potential of the ordinary KS equation.  There is both a difference of
content \cite{YangPengSunPerdew2016,PerdewEtAlBandGaps2017}
and a computational cost penalty for gKS compared to KS.  We had
addressed both those issues by deorbitalization, that is, the replacement
of the orbital dependence with a function of the density, its gradient, 
and its Laplacian. That gave the SCAN-L DFA \cite{SCANL1,SCANL2}.
Except for elemental 3$d$ magnetic solids, SCAN-L delivered essentially
the same performance as SCAN.  It should be faster than SCAN but in
practice numerical instabilities caused very slow SCF convergence.
Very recently we found that the greatly improved numerical stability
of $\rrscan$ is preserved under deorbitalization to yield $\rrscan$-L.
In solid calculations, it runs almost 4 times faster than $\rrscan$
\cite{rrSCANL}.

The advent of $\rrscan$ and $\rrscan$-L make it imperative to investigate
their SCO performance on the Ref.\ \citenum{CireraNadalRuiz2018} data set to see if the changes
from SCAN and SCAN-L (respectively) affect the delicate energy differences
involved.

We note that validation and testing of any DFA with regard to SCO
must be preceded by standard screening.  Any plausible DFA candidate for SCO 
first must have given acceptable
accuracy for molecular dissociation energies, bond lengths, and
fundamental vibrational frequencies, and must give acceptable
crystal structures, cohesive energies, and bulk moduli, all against
widely used data bases.  If the fundamental molecular
vibration frequencies and bulk moduli are of decent quality, it is
plausible that the enthalpic contributions to the SCO temperature
$T_\half$ will be reasonable. Notice that the \emph{essential} untested
ingredient from routine DFA screening is $\Delta E_{HL}$.  If the
candidate DFA delivers bad values for that difference, the only way it could deliver
good $T_\half$ values would be by compensating error, i.e., right answers
for wrong reasons. We focus therefore on $\Delta E_{HL}$.  For the sake
of delineating DFA performance in difficult spin systems, we also study
the Cr$_2$ dissociation curve. 

\section{Computational Methods}

Molecular calculations were done with a locally modified developers' version 
of the NWChem code \cite{NWChem2020} using the unrestricted KS procedure,
 the \texttt{def2-TZVP} basis set \cite{def2} in
spherical representation, and the \texttt{FINE} numerical integration grid. 
The number of radial shells and the corresponding Lebedev angular
points \cite{Lebedev75,Lebedev76,Lebedev77,LebedevSkorokhodov,Lebedev94,LebedevLaikov} 
per radial shell vary depending on the atom type, as shown in Table \ref{grid}.
Previously we have shown that this grid density is good enough to integrate
both \rrscan\ and \rrscan-L XC potentials and energies \cite{rrSCANL}.
SCAN calculations used a custom-defined grid with 200 radial shells 
and 590 angular points per shell.
\begin{table}
\caption{Number of radial and Lebedev angular points present in
the NWChem \texttt{FINE} grid preset \label{grid}}
\begin{tabular}{l c c }
Element            & Radial shells & Angular points \\\hline\hline
H                  &  60   & 590 \\
B, C, N, O         &  70   & 590 \\
P, S               &  123  & 770 \\
Cr, Mn, Fe, Co, Br &  140  & 974 \\
I                  &  141  & 974 \\\hline\hline
\end{tabular}
\end{table}
Moreover, all calculations used Weigend's Coulomb-fitting basis set
\cite{Weigend2006} 
for the density fitting scheme \cite{Whitten1973,DunlapConollySabin1979}. 
Default options for guess density, convergence stabilization and
acceleration techniques, and convergence criteria for both 
electronic and ionic relaxations, were used.

The D3(BJ)\cite{DFTD3,DFTD3BJ}
empirical dispersion corrections, with parameters optimized
for SCAN \cite{BrandenburgEtAl16}, was tried as an exploratory step.
We remark that both \rrscan\ and \rrscan-L should include
some mid-range dispersion by construction, so the D3(BJ) corrections
are rather small.

Nine SCO systems in the data-base are
positively charged (2 Mn$^{\mathrm{III}}$, 3 Fe$^{\mathrm{III}}$, 
1 Fe$^{\mathrm{II}}$, and 3 Co$^{\mathrm{II}}$). None
of the counter-ions were included in the calculations. That
corresponds to removal of  10 \% or less of the total atomic count
for most of the charged complexes. However, the 45 atoms of the
tetraphenyl borate anion originally present in the Fe$^{\mathrm{III}}$ system
labeled S9, account for almost 40 \% of the total number of atoms of that system.
 
The spin-state energetics for the four Fe complexes, three 
Fe$^{\mathrm{III}}$ and one Fe$^{\mathrm{II}}$,
recently benchmarked by Rado\'n \cite{Radon2019} also were computed. 
Geometries of all four were re-optimized under the same symmetry
constraints as in Ref. \citenum{Radon2019} using the same settings 
as for systems in the  Ref.\ \citenum{CireraNadalRuiz2018}  data-base.

The chromium dimer potential energy curves were obtained using 
a modified version of VASP 5.4.4 \cite{vasp,vasp2,vasp3,vasp4} using the
14-electron projector augmented-wave (PAW) dataset\cite{paw1,paw2}. 
The dimer was aligned along 
the $z$-axis inside a large $12\times12\times 15$ $\AA^3$ box.
The calculations used the \emph{accurate} precision setting,
a 600 eV kinetic energy cutoff, and included aspherical corrections
inside the PAW spheres.

\section{Results and Discussion}
\begin{figure}
\includegraphics[width=0.45\textwidth]{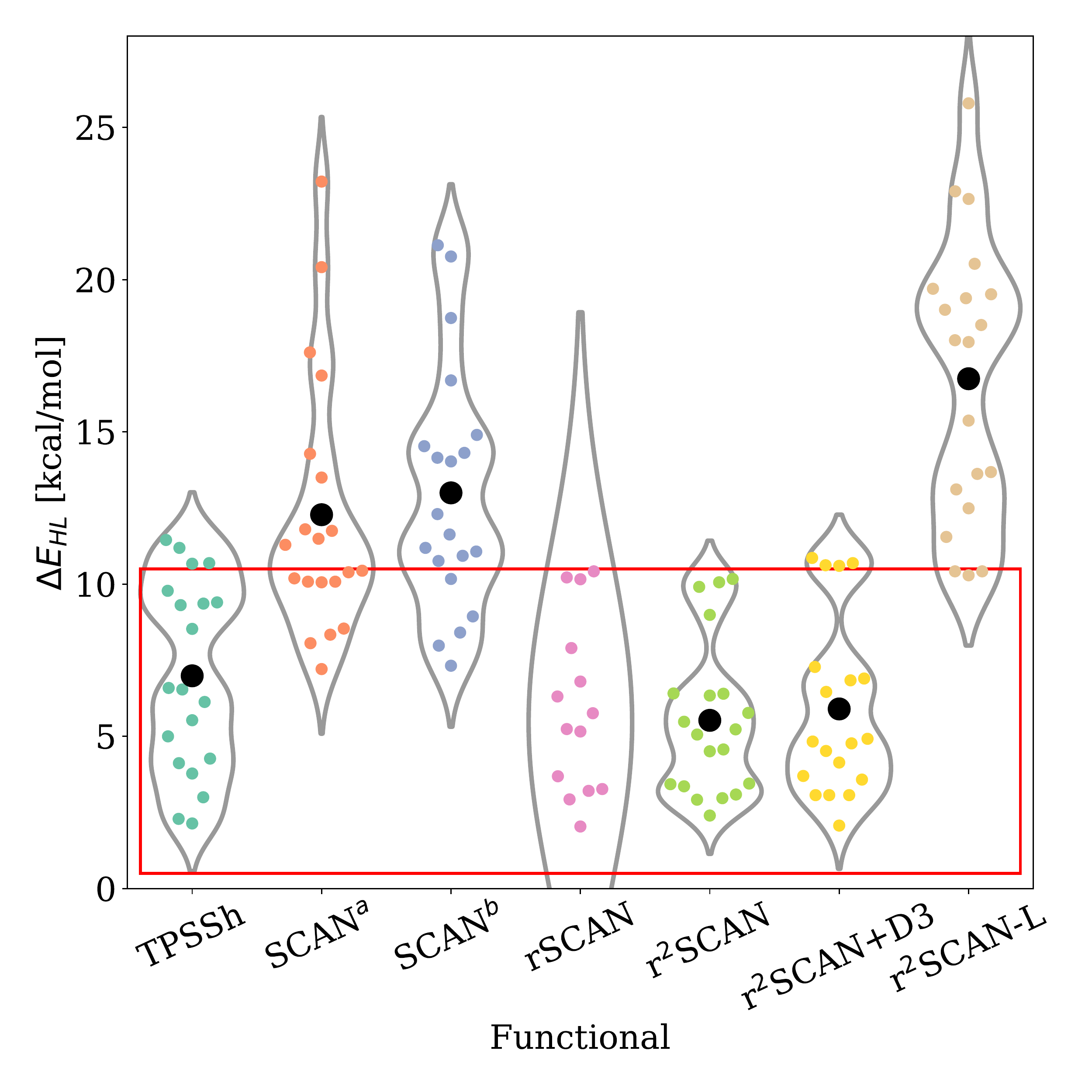}
\caption{Electronic low-spin-high-spin energy differences $\Delta E_{HL}$ in kcal/mol.
The colored dots correspond to the actual individual results, while the large black
dot corresponds to the mean $\Delta E_{HL}$ obtained with each functional. The red box
is the same as used in Ref. \citenum{CireraRuiz2020} to indicate the region where the electronic energy
difference can be compensated by the entropy in usual SCO systems. Note that the
rSCAN violin was cut in order to enhance visibility of other results. 
TPSSh and SCAN$^a$ from Ref. \citenum{CireraRuiz2020}. All other this work. \label{figure}}
\end{figure}

Table \ref{table} and Fig. \ref{figure} show $\Delta E_{HL}$,
in kcal/mol, obtained with SCAN, rSCAN, \rrscan, and \rrscan-L. Results from
Ref. \citenum{CireraRuiz2020} for TPSSh and SCAN are included
for comparison. SCAN results correspond
to the denser numerical integration grid (SG-2) which
is made up of 75 radial shells with 302 Lebedev angular points
per shell. Our $\Delta E_{HL}$ values obtained with SCAN are
around 1.5 kcal/mol from those of Ref. \citenum{CireraRuiz2020}.
This difference was expected based on our previous
studies (see Refs. \citenum{rSCANcomment} and \citenum{rrSCANL}).

Reassuringly, 
\rrscan\ gives the correct low-spin configuration for the
ground-state for all 20 complexes, with all $\Delta E_{HL}$
values inside a 10 kcal/mol energy window proposed by Cirera and Ruiz
(compare Fig. \ref{figure} with Fig. 3 of Ref. \citenum{CireraRuiz2020}).
Also striking is the fact that the \rrscan\ DFA yields a marked 
reduction of the predicted $\Delta E_{HL}$ 
compared to SCAN. The \rrscan\ values are, in fact,  
slightly smaller than those predicted by the DFA hybrid TPSSh. 

In contrast, the deorbitalized version, \rrscan-L, gives 
$\Delta E_{HL}$ larger than, but still comparable to, 
the values obtained with SCAN. The advantage of \rrscan-L 
is mainly the potential speed-up one can achieve
by means of its local multiplicative potential. We
return to this point below.

Table \ref{table} also shows that the inclusion of 
empirical dispersion corrections via
the DFT-D3 approach \cite{DFTD3,DFTD3BJ} changes 
$\Delta E_{HL}$ values by about 0.5 kcal/mol, with
larger effects for the Co systems. We stress
that forces from the \rrscan+D3 combination were included during
geometry optimizations.  We did not find
unrealistic geometries such as reported in 
Reference \citenum{CireraNadalRuiz2018}. Larger
effects generally are seen when the DFT-D3 correction
is used only for single-point energies at the
corresponding uncorrected DFA minima 
(see for example Ref. \citenum{CireraRuiz2020}.)

In order to obtain further insight about the specific changes 
that lead to the drastic performance differences among these closely
related DFAs, we also tried rSCAN calculations.
As Fig. \ref{figure} and Table \ref{table} show, rSCAN 
does almost as well as \rrscan\ for the majority of systems, 
but gives the wrong sign for six of them. The most notable 
failures occur in Fe$^{\mathrm{II}}$ $d^6$ systems (S11-S15). In them, 
rSCAN overstabilizes the high-spin state by as much as
54 kcal/mol. 

It is interesting to note that, barring
rSCAN results for Fe(II) systems, the S9 $\Delta E_{HL}$
is the largest of the set for the TPSSh, \rrscan\ ,
\rrscan-L, and rSCAN DFAs. This may be a direct consequence
of the effects that the missing counter-ion can have on the 
overall structure and energetics of the complex 
(see Computational Methods), but investigation of 
the issue is outside the scope of this
work.

\begin{table}
\caption{High-spin to low-spin $\Delta E_{HL}$ energies [kcal/mol]. 
Systems are labeled as in Cirera and Ruiz \cite{CireraRuiz2020}.
\label{table}}
\begin{tabular}{l r r r r r r r }
System & TPSSh\textsuperscript{\emph{a}} & SCAN\textsuperscript{\emph{a}} & SCAN\textsuperscript{\emph{b}} &
 rSCAN & \rrscan & \rrscan+D3 & \rrscan-L  \\\hline\hline
S1     & 6.54     & 11.49   & 7.98  &  2.93  &  3.36    & 4.14   &  10.54 \\
S2     & 4.27     & 8.06    & 8.94  &  3.27  &  3.09    & 3.07   &  11.55 \\
S3     & 5.53     & 8.54    & 8.41  &  3.69  &  3.45    & 3.58   &  19.39 \\
S4     & 4.12     & 7.21    & 7.32  &  2.04  &  2.40    & 2.07   &  10.42 \\
S5     & 11.19    & 10.08   & 10.93 & -2.27  &  5.06    & 4.92   &  12.49 \\
S6     & 10.67    & 10.39   & 11.07 &  5.24  &  4.57    & 4.77   &  19.52 \\
S7     & 9.40     & 10.19   & 11.19 &  5.16  &  4.51    & 4.83   &  19.70 \\
S8     & 9.78     & 11.29   & 11.63 &  3.21  &  2.92    & 3.07   &  18.51 \\
S9     & 11.45    & 17.61   & 18.74 & 10.42  & 10.17    & 10.62  &  25.79 \\
S10    & 10.69    & 14.28   & 14.90 &  5.76  &  5.48    & 6.02   &  20.52 \\
S11    &  6.13    & 13.50   & 14.31 & -26.43 &  5.23    & 4.52   &  18.01 \\
S12    &  8.53    & 16.85   & 16.69 & -54.07 &  6.34    & 6.46   &  19.01 \\
S13    &  9.31    & 20.41   & 20.76 & -15.64 & 10.06    & 10.86  &  22.65 \\
S14    &  9.36    & 23.22   & 21.13 & -19.95 &  9.91    & 10.60  &  22.90 \\
S15    &  5.00    & 11.75   & 14.01 & -28.96 &  2.97    & 3.07   &  15.37 \\
S16    &  3.00    & 10.44   & 12.30 &  7.90  &  6.40    & 6.84   &  13.68 \\
S17    &  2.29    & 8.34    & 14.53 &  10.16 &  3.43    & 3.70   &  10.29 \\
S18    &  2.14    & 10.08   & 10.17 &   6.31 &  5.77    & 7.28   &  13.11 \\
S19    &  3.78    & 11.80   & 14.15 &  10.22 &  8.99    & 10.69  &  17.95 \\
S20    &  6.59    & 10.06   & 10.76 &   6.80 &  6.41    & 6.90   &  13.62 \\\hline\hline
\end{tabular}\\
\textsuperscript{\emph{a}} From Ref. \citenum{CireraRuiz2020}.
\textsuperscript{\emph{b}} This work.
\end{table}

For thoroughness, we augmented the 20-system data-set with the four
systems proposed by Rado\'n in Ref. \citenum{Radon2019}. Rado\'n gave 
$\Delta E_{HL}$ values by removing environmental effects
from experimental data in order to have a straightforward
reference for comparison.  
Table \ref{radon} shows  $\Delta E_{HL}$ values for those
systems (labeled as S21--S24) and the corresponding 
reference reported as extrapolated from experiment. 
Noe that system S24 is the same as S8 reported of Table \ref{table}.
The TPSSh and TPSSh+D3 values 
taken from Ref. \citenum{Radon2019} are from non-relativistic 
calculations without corrections for spin-contamination.
Also note that $\Delta E_{HL}$ for  S21 and S22
corresponds to vertical excitation energies, so the DFT-D3 corrected and
uncorrected values are exactly the same. Again, we see that
\rrscan\ gives the correct sign and outperforms TPSSh(+D3).

Based on the foregoing results, we can summarize our
findings as follows:

\begin{itemize}
\item The reduced $\Delta E_{HL}$ values (compared to those from SCAN)  obtained
with rSCAN and \rrscan\ mean that the seemingly small changes made 
in the SCAN switching function are responsible for the 
majority of the effects.
\item The failures obtained from rSCAN but not  with
\rrscan\ highlight the importance that constraint satisfaction
has in ensuring maximum scope of validity for a given DFA.
\item The larger $\Delta E_{HL}$ values obtained with
SCAN-L and \rrscan-L also illuminate the importance of the
switching function in the prediction of spin-state energetics. Although
the de-orbitalization procedure \cite{SCANL1,SCANL2} does not 
change the switching function \emph{directly}, the differences between 
the approximated iso-orbital indicator 
$\alpha_L$ and the original one modify, indirectly, its
behavior \cite{DMR-SBT-SCANmag}.
\end{itemize}

\begin{table}
\caption{High-spin to low-spin $\Delta E_{HL}$ energies [kcal/mol] of complexes S21--S24.\label{radon}}
\begin{tabular}{l c c c c}
                                    & S21  & S22  &  S23  & S24 (S8) \\\hline\hline
TPSSh\textsuperscript{\emph{a}}     & -29.2 & 24.7 &  7.6 & 10.8     \\
TPSSh+D3\textsuperscript{\emph{a}}  & -29.2 & 24.7 &  8.3 & 10.2     \\
\rrscan                             & -45.4 & 22.4 &  6.8 & 2.9      \\
Exptl.\textsuperscript{\emph{a}}    & -47.4 & 19.7 &  3.8 & 2.4      \\\hline\hline
\end{tabular}

\textsuperscript{\emph{a}} Non-relativistic values from Ref. \citenum{Radon2019}. 
\end{table}

\section{Conclusions and Outlook}
We have shown that the \rrscan\ DFA provides
a quantitatively correct ground state for all 
molecules in the SCO data-base put forth in Ref. 
\citenum{CireraNadalRuiz2018} as well as on the four
Fe complexes in Ref.\ \citenum{Radon2019}. Furthermore,
\rrscan\ apparently is the only comparatively
simple DFA, including hybrid ones,
that gives all high-spin to low-spin energy
differences $\Delta_{HL}$ inside what is 
believed to be the appropriate energy range.
On the basis of that accuracy and comparatively
modest computational costs, we therefore 
recommend, strongly, the use of \rrscan\ to describe 3$d$
SCO systems. 

Though the accuracy for $\Delta E_{HL}$ provided
by the deorbitalized version, \rrscan-L, is not
as good as what \rrscan\ gives, it is useful 
that \rrscan-L does perform on par with the 
accuracy from original SCAN, but without the
numerical integration issues. The advantage of
\rrscan-L is its substantially lower computational
costs in codes that use fast-Fourier transforms
to obtain the appropriate derivatives of the
density. In those codes, the local multiplicative potential of
\rrscan-L can achieve calculations as much as
4 times faster than with \rrscan. That provides a
major opportunity. The key to it is that the sacrifice in bond length 
and vibrational frequency accuracy in going from
\rrscan\ to \rrscan-L is much smaller than the 
$\Delta E_{HL}$ accuracy sacrifice. The strategy we recommend
therefore is to do geometry optimizations (either molecular
or solid) with \rrscan-L, then do a single point calculation
with \rrscan\ to determine $\Delta E_{HL}$. We have 
that strategy under investigation, but would like
to point out that an analogous approach, using SCAN and SCAN-L,
has proven to be successful \cite{HinzEtAl2020}.

\begin{figure}
\includegraphics[width=0.45\textwidth]{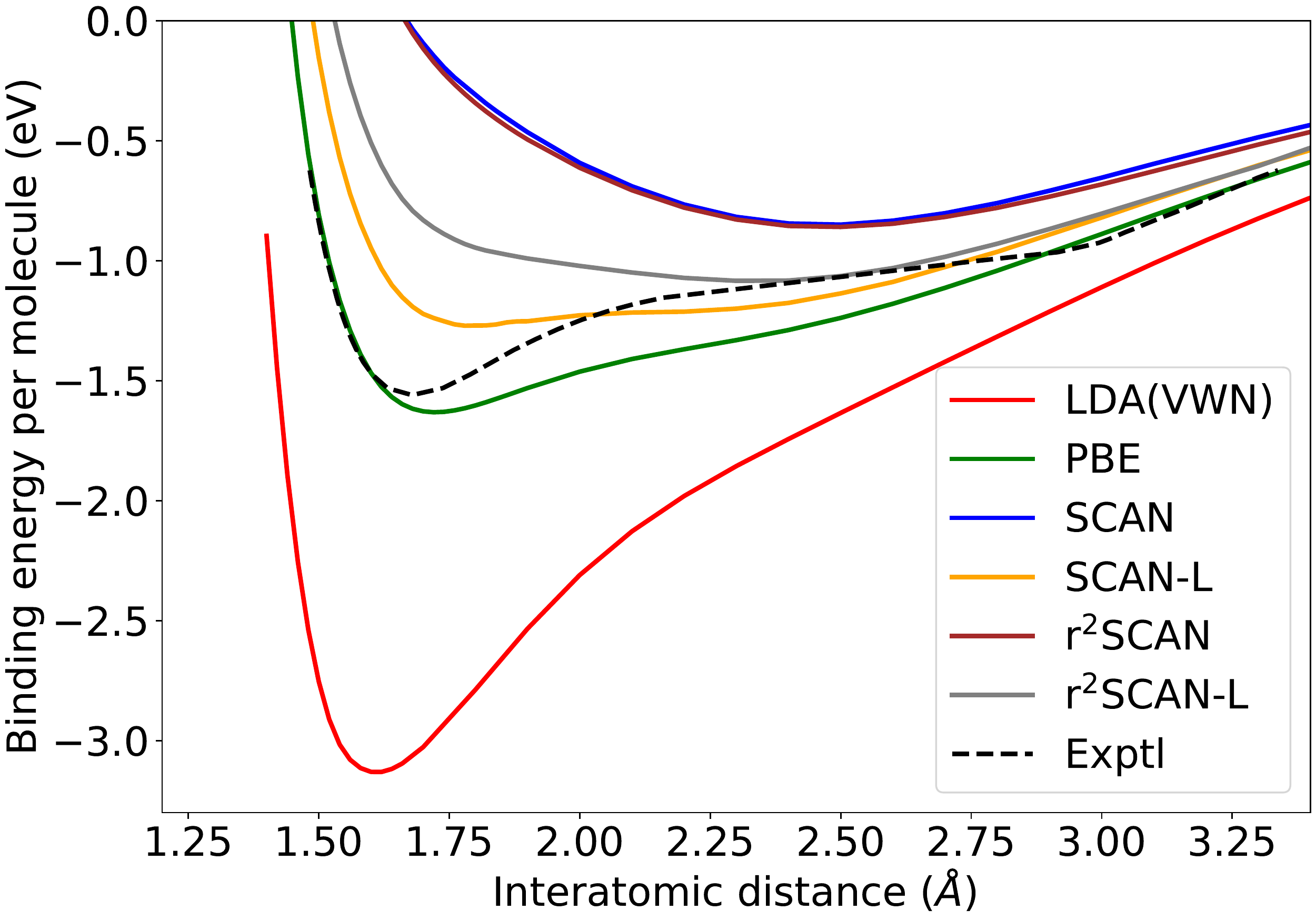}
\caption{Cr$_2$ potential energy curve [eV] obtained with six different
DFAs in VASP. The experimental curve is reproduced from 
Ref. \citenum{CaseyLeopold1993} assuming that the bottom of the well
is at -1.56 eV \cite{SimardEtAl1998,Vanco2016}.  \label{cr2}}
\end{figure}

While \rrscan\ is much better for SCO on the Cirera-Nadal-Ruiz,
and Rado\'n data sets
and therefore we recommend it, we do so with caution. \rrscan\ is not 
perfect for magnetization nor is \rrscan-L.  See Fig. \ref{cr2} for a 
comparison of \rrscan\ and \rrscan-L results with those from other DFAs 
for the famously difficult case of Cr$_2$ dissociation, an issue 
recently re-addressed in Ref. \citenum{ZhangZhangSingh}  (Note that 
the corresponding figure in Ref. \citenum{ZhangZhangSingh} displays the 
experimental data incorrectly by a factor of 2.)  The reasons for the 
superior performance of the GGA functional PBE compared to any of the 
meta-GGA functionals on this system remain obscure to us and to 
other DFA developers\cite{PerdewPC}. Another aspect of the obscurity is that,
distinct from SCAN, \rrscan\ and both de-orbitalized functionals do not
magnetize benzene or graphene.  This slight improvement is also seen
in the solid phase, since both \rrscan\ and \rrscan-L improve upon SCAN 
(see Table III of Ref. \citenum{rrSCANL}.)
Caution is still warranted in use of \rrscan\
and \rrscan-L according to the protocol we have proposed here.

\begin{suppinfo}
Chemical formulas and \rrscan\ optimized structures
for the 23 complexes studied.
\end{suppinfo}

\begin{acknowledgement}
This work was supported by 
U.S.\ Dept. of Energy under Energy Frontier Research Center 
grant DE-SC 0019330.
\end{acknowledgement}

\bibliography{SCORefs}

\providecommand{\latin}[1]{#1}
\makeatletter
\providecommand{\doi}
  {\begingroup\let\do\@makeother\dospecials
  \catcode`\{=1 \catcode`\}=2 \doi@aux}
\providecommand{\doi@aux}[1]{\endgroup\texttt{#1}}
\makeatother
\providecommand*\mcitethebibliography{\thebibliography}
\csname @ifundefined\endcsname{endmcitethebibliography}
  {\let\endmcitethebibliography\endthebibliography}{}
\begin{mcitethebibliography}{54}
\providecommand*\natexlab[1]{#1}
\providecommand*\mciteSetBstSublistMode[1]{}
\providecommand*\mciteSetBstMaxWidthForm[2]{}
\providecommand*\mciteBstWouldAddEndPuncttrue
  {\def\EndOfBibitem{\unskip.}}
\providecommand*\mciteBstWouldAddEndPunctfalse
  {\let\EndOfBibitem\relax}
\providecommand*\mciteSetBstMidEndSepPunct[3]{}
\providecommand*\mciteSetBstSublistLabelBeginEnd[3]{}
\providecommand*\EndOfBibitem{}
\mciteSetBstSublistMode{f}
\mciteSetBstMaxWidthForm{subitem}{(\alph{mcitesubitemcount})}
\mciteSetBstSublistLabelBeginEnd
  {\mcitemaxwidthsubitemform\space}
  {\relax}
  {\relax}

\bibitem[Ioannidis and Kulik(2015)Ioannidis, and Kulik]{IoannidisKulik2015}
Ioannidis,~E.~I.; Kulik,~H.~J. {Towards quantifying the role of exact exchange
  in predictions of transition metal complex properties}. \emph{J. Chem. Phys.}
  \textbf{2015}, \emph{143}, 34104\relax
\mciteBstWouldAddEndPuncttrue
\mciteSetBstMidEndSepPunct{\mcitedefaultmidpunct}
{\mcitedefaultendpunct}{\mcitedefaultseppunct}\relax
\EndOfBibitem
\bibitem[Mortensen and Kepp(2015)Mortensen, and Kepp]{MortensenKepp2015}
Mortensen,~S.~R.; Kepp,~K.~P. {Spin Propensities of Octahedral Complexes From
  Density Functional Theory}. \emph{J. Phys. Chem. A} \textbf{2015},
  \emph{119}, 4041--4050\relax
\mciteBstWouldAddEndPuncttrue
\mciteSetBstMidEndSepPunct{\mcitedefaultmidpunct}
{\mcitedefaultendpunct}{\mcitedefaultseppunct}\relax
\EndOfBibitem
\bibitem[Cirera and Ruiz(2016)Cirera, and Ruiz]{CireraRuiz2016}
Cirera,~J.; Ruiz,~E. {Theoretical Modeling of the Ligand-Tuning Effect over the
  Transition Temperature in Four-Coordinated Fe$^{\mathrm{II}}$ Molecules}.
  \emph{Inorg. Chem.} \textbf{2016}, \emph{55}, 1657--1663\relax
\mciteBstWouldAddEndPuncttrue
\mciteSetBstMidEndSepPunct{\mcitedefaultmidpunct}
{\mcitedefaultendpunct}{\mcitedefaultseppunct}\relax
\EndOfBibitem
\bibitem[Harding \latin{et~al.}(2016)Harding, Harding, and
  Phonsri]{HardingHardingPhonsri2016}
Harding,~D.~J.; Harding,~P.; Phonsri,~W. {Spin crossover in iron(III)
  complexes}. \emph{Coord. Chem. Rev.} \textbf{2016}, \emph{313}, 38--61\relax
\mciteBstWouldAddEndPuncttrue
\mciteSetBstMidEndSepPunct{\mcitedefaultmidpunct}
{\mcitedefaultendpunct}{\mcitedefaultseppunct}\relax
\EndOfBibitem
\bibitem[Kepp(2016)]{Kepp2016}
Kepp,~K.~P. {Theoretical Study of Spin Crossover in 30 Iron Complexes}.
  \emph{Inorg. Chem.} \textbf{2016}, \emph{55}, 2717--2727\relax
\mciteBstWouldAddEndPuncttrue
\mciteSetBstMidEndSepPunct{\mcitedefaultmidpunct}
{\mcitedefaultendpunct}{\mcitedefaultseppunct}\relax
\EndOfBibitem
\bibitem[Amabilino and Deeth(2017)Amabilino, and Deeth]{AmalbinoDeeth2017}
Amabilino,~S.; Deeth,~R.~J. {DFT Analysis of Spin Crossover in Mn(III)
  Complexes: Is a Two-Electron $S=2$ to $S=0$ Spin Transition Feasible?}
  \emph{Inorg. Chem.} \textbf{2017}, \emph{56}, 2602--2613\relax
\mciteBstWouldAddEndPuncttrue
\mciteSetBstMidEndSepPunct{\mcitedefaultmidpunct}
{\mcitedefaultendpunct}{\mcitedefaultseppunct}\relax
\EndOfBibitem
\bibitem[Flores-Leonar \latin{et~al.}(2017)Flores-Leonar, Moreno-Esparza,
  Ugalde-Saldıvar, and Amador-Bedolla]{Flores-LeonarEtAl2017}
Flores-Leonar,~M.~M.; Moreno-Esparza,~R.; Ugalde-Saldıvar,~V.~M.;
  Amador-Bedolla,~C. {Correlating Properties in Iron(III) Complexes: A DFT
  Description of Structure, Redox Potential and Spin Crossover Phenomena}.
  \emph{ChemistrySelect} \textbf{2017}, \emph{2}, 4717--4724\relax
\mciteBstWouldAddEndPuncttrue
\mciteSetBstMidEndSepPunct{\mcitedefaultmidpunct}
{\mcitedefaultendpunct}{\mcitedefaultseppunct}\relax
\EndOfBibitem
\bibitem[Sirirak \latin{et~al.}(2017)Sirirak, Sertphon, Phonsri, Harding, and
  Harding]{SirirakEtAl2017}
Sirirak,~J.; Sertphon,~D.; Phonsri,~W.; Harding,~P.; Harding,~D.~J. {Comparison
  of density functionals for the study of the high spin low spin gap in Fe(III)
  spin crossover complexes}. \emph{Int. J. Quantum Chem.} \textbf{2017},
  \emph{117}, e25362\relax
\mciteBstWouldAddEndPuncttrue
\mciteSetBstMidEndSepPunct{\mcitedefaultmidpunct}
{\mcitedefaultendpunct}{\mcitedefaultseppunct}\relax
\EndOfBibitem
\bibitem[Cirera \latin{et~al.}(2018)Cirera, Via-Nadal, and
  Ruiz]{CireraNadalRuiz2018}
Cirera,~J.; Via-Nadal,~M.; Ruiz,~E. {Benchmarking Density Functional Methods
  for Calculation of State Energies of First Row Spin-Crossover Molecules}.
  \emph{Inorg. Chem.} \textbf{2018}, \emph{57}, 14097--14105\relax
\mciteBstWouldAddEndPuncttrue
\mciteSetBstMidEndSepPunct{\mcitedefaultmidpunct}
{\mcitedefaultendpunct}{\mcitedefaultseppunct}\relax
\EndOfBibitem
\bibitem[Staroverov \latin{et~al.}(2003)Staroverov, Scuseria, Tao, and
  Perdew]{TPSSh}
Staroverov,~V.~N.; Scuseria,~G.~E.; Tao,~J.; Perdew,~J.~P. {Comparative
  assessment of a new nonempirical density functional: Molecules and
  hydrogen-bonded complexes}. \emph{J. Chem. Phys.} \textbf{2003}, \emph{119},
  12129\relax
\mciteBstWouldAddEndPuncttrue
\mciteSetBstMidEndSepPunct{\mcitedefaultmidpunct}
{\mcitedefaultendpunct}{\mcitedefaultseppunct}\relax
\EndOfBibitem
\bibitem[Tao \latin{et~al.}(2003)Tao, Perdew, Staroverov, and Scuseria]{TPSS}
Tao,~J.; Perdew,~J.~P.; Staroverov,~V.~N.; Scuseria,~G.~E. {Climbing the
  Density Functional Ladder: Nonempirical Meta-Generalized Gradient
  Approximation Designed for Molecules and Solids}. \emph{Phys. Rev. Lett.}
  \textbf{2003}, \emph{91}, 146401\relax
\mciteBstWouldAddEndPuncttrue
\mciteSetBstMidEndSepPunct{\mcitedefaultmidpunct}
{\mcitedefaultendpunct}{\mcitedefaultseppunct}\relax
\EndOfBibitem
\bibitem[Perdew and Schmidt(2001)Perdew, and Schmidt]{PerdewSchmidt2001}
Perdew,~J.~P.; Schmidt,~K. {Jacob's ladder of density functional approximations
  for the exchange-correlation energy}. \emph{AIP Conf. Proc.} \textbf{2001},
  \emph{577}, 1\relax
\mciteBstWouldAddEndPuncttrue
\mciteSetBstMidEndSepPunct{\mcitedefaultmidpunct}
{\mcitedefaultendpunct}{\mcitedefaultseppunct}\relax
\EndOfBibitem
\bibitem[Sun \latin{et~al.}(2015)Sun, Ruzsinszky, and Perdew]{SCAN}
Sun,~J.; Ruzsinszky,~A.; Perdew,~J.~P. {Strongly Constrained and Appropriately
  Normed Semilocal Density Functional}. \emph{Phys. Rev. Lett.} \textbf{2015},
  \emph{115}, 36402\relax
\mciteBstWouldAddEndPuncttrue
\mciteSetBstMidEndSepPunct{\mcitedefaultmidpunct}
{\mcitedefaultendpunct}{\mcitedefaultseppunct}\relax
\EndOfBibitem
\bibitem[Sun \latin{et~al.}(2016)Sun, Remsing, Zhang, Sun, Ruzsinszky, Peng,
  Yang, Pau, Waghmare, Wu, Klein, and Perdew]{SCANNature}
Sun,~J.; Remsing,~R.~C.; Zhang,~Y.; Sun,~Z.; Ruzsinszky,~A.; Peng,~H.;
  Yang,~Z.; Pau,~A.; Waghmare,~U.; Wu,~X. \latin{et~al.}  {Accurate
  first-principles structures and energies of diversely bonded systems from an
  efficient density functional}. \emph{Nature Chemistry} \textbf{2016},
  \emph{8}, 831--836\relax
\mciteBstWouldAddEndPuncttrue
\mciteSetBstMidEndSepPunct{\mcitedefaultmidpunct}
{\mcitedefaultendpunct}{\mcitedefaultseppunct}\relax
\EndOfBibitem
\bibitem[Mej{\'{i}}a-Rodr{\'{i}}guez and
  Trickey(2019)Mej{\'{i}}a-Rodr{\'{i}}guez, and Trickey]{DMR-SBT-SCANmag}
Mej{\'{i}}a-Rodr{\'{i}}guez,~D.; Trickey,~S.~B. {Analysis of Over-magnetization
  of Elemental Transition Metal Solids from the SCAN Density Functional}.
  \emph{Phys. Rev. B} \textbf{2019}, \emph{100}, 041113(R)\relax
\mciteBstWouldAddEndPuncttrue
\mciteSetBstMidEndSepPunct{\mcitedefaultmidpunct}
{\mcitedefaultendpunct}{\mcitedefaultseppunct}\relax
\EndOfBibitem
\bibitem[Cirera and Ruiz(2020)Cirera, and Ruiz]{CireraRuiz2020}
Cirera,~J.; Ruiz,~E. {Assessment of the SCAN Functional for Spin State Energies
  in Spin-Crossover Systems}. \emph{J. Phys. Chem. A} \textbf{2020},
  \emph{124}, 5053--5058\relax
\mciteBstWouldAddEndPuncttrue
\mciteSetBstMidEndSepPunct{\mcitedefaultmidpunct}
{\mcitedefaultendpunct}{\mcitedefaultseppunct}\relax
\EndOfBibitem
\bibitem[Bart\'ok and Yates(2019)Bart\'ok, and Yates]{rSCAN}
Bart\'ok,~A.~P.; Yates,~J.~R. {Regularized SCAN functional}. \emph{Journal of
  Chemical Physics} \textbf{2019}, \emph{150}, 161101\relax
\mciteBstWouldAddEndPuncttrue
\mciteSetBstMidEndSepPunct{\mcitedefaultmidpunct}
{\mcitedefaultendpunct}{\mcitedefaultseppunct}\relax
\EndOfBibitem
\bibitem[Mej{\'{i}}a-Rodr{\'{i}}guez and
  Trickey(2019)Mej{\'{i}}a-Rodr{\'{i}}guez, and Trickey]{rSCANcomment}
Mej{\'{i}}a-Rodr{\'{i}}guez,~D.; Trickey,~S.~B. {Comment on "Regularized SCAN
  functional" [J. Chem. Phys. \textbf{150}, 161101 (2019)]}. \emph{J. Chem.
  Phys.} \textbf{2019}, \emph{151}, 207101\relax
\mciteBstWouldAddEndPuncttrue
\mciteSetBstMidEndSepPunct{\mcitedefaultmidpunct}
{\mcitedefaultendpunct}{\mcitedefaultseppunct}\relax
\EndOfBibitem
\bibitem[Peng \latin{et~al.}(2016)Peng, Yang, Perdew, and Sun]{SCANrVV10}
Peng,~H.; Yang,~Z.-H.; Perdew,~J.~P.; Sun,~J. {Versatile van der Waals Density
  Functional based on a Meta-Generalized Gradient Approximation}. \emph{Phys.
  Rev. X} \textbf{2016}, 41005\relax
\mciteBstWouldAddEndPuncttrue
\mciteSetBstMidEndSepPunct{\mcitedefaultmidpunct}
{\mcitedefaultendpunct}{\mcitedefaultseppunct}\relax
\EndOfBibitem
\bibitem[Tran \latin{et~al.}(2016)Tran, Stelzl, and Blaha]{TranStelzlBlaha}
Tran,~F.; Stelzl,~J.; Blaha,~P. {Rungs 1 to 4 of DFT Jacob’s ladder:
  Extensive test on the lattice constant, bulk modulus, and cohesive energy of
  solids}. \emph{J. Chem. Phys.} \textbf{2016}, \emph{144}, 204120\relax
\mciteBstWouldAddEndPuncttrue
\mciteSetBstMidEndSepPunct{\mcitedefaultmidpunct}
{\mcitedefaultendpunct}{\mcitedefaultseppunct}\relax
\EndOfBibitem
\bibitem[Furness \latin{et~al.}(2020)Furness, Kaplan, Ning, Perdew, and
  Sun]{FurnessEtAl2020}
Furness,~J.~W.; Kaplan,~A.~D.; Ning,~J.; Perdew,~J.~P.; Sun,~J. {Accurate and
  numerically efficient r$^2$SCAN meta-generalized gradient approximation}.
  \emph{J. Phys. Chem. Lett.} \textbf{2020}, \emph{11}, 8208--8215\relax
\mciteBstWouldAddEndPuncttrue
\mciteSetBstMidEndSepPunct{\mcitedefaultmidpunct}
{\mcitedefaultendpunct}{\mcitedefaultseppunct}\relax
\EndOfBibitem
\bibitem[Yang \latin{et~al.}(2016)Yang, Peng, Sun, and
  Perdew]{YangPengSunPerdew2016}
Yang,~Z.-h.; Peng,~H.; Sun,~J.; Perdew,~J.~P. {More realistic band gaps from
  meta-generalized gradient approximations: Only in a generalized Kohn-Sham
  scheme}. \emph{Phys. Rev. B} \textbf{2016}, \emph{93}, 205205\relax
\mciteBstWouldAddEndPuncttrue
\mciteSetBstMidEndSepPunct{\mcitedefaultmidpunct}
{\mcitedefaultendpunct}{\mcitedefaultseppunct}\relax
\EndOfBibitem
\bibitem[Perdew \latin{et~al.}(2017)Perdew, Yang, Burke, Yang, Gross,
  Scheffler, Scuseria, Henderson, Zhang, Ruzsinszky, Peng, Sun, Trushin, and
  G{\"o}rling]{PerdewEtAlBandGaps2017}
Perdew,~J.; Yang,~W.; Burke,~K.; Yang,~Z.; Gross,~E.; Scheffler,~M.;
  Scuseria,~G.; Henderson,~T.; Zhang,~I.; Ruzsinszky,~A. \latin{et~al.}
  {Understanding band gaps of solids in generalized Kohn–Sham theory}.
  \emph{Proc. Natl. Acad. Sci. (USA)} \textbf{2017}, \emph{114}, 2801\relax
\mciteBstWouldAddEndPuncttrue
\mciteSetBstMidEndSepPunct{\mcitedefaultmidpunct}
{\mcitedefaultendpunct}{\mcitedefaultseppunct}\relax
\EndOfBibitem
\bibitem[Mej{\'{i}}a-Rodr{\'{i}}guez and
  Trickey(2017)Mej{\'{i}}a-Rodr{\'{i}}guez, and Trickey]{SCANL1}
Mej{\'{i}}a-Rodr{\'{i}}guez,~D.; Trickey,~S.~B. {Deorbitalization strategies
  for meta-generalized-gradient-approximation exchange-correlation
  functionals}. \emph{Phys. Rev. A} \textbf{2017}, \emph{96}, 052512\relax
\mciteBstWouldAddEndPuncttrue
\mciteSetBstMidEndSepPunct{\mcitedefaultmidpunct}
{\mcitedefaultendpunct}{\mcitedefaultseppunct}\relax
\EndOfBibitem
\bibitem[Mej{\'{i}}a-Rodr{\'{i}}guez and
  Trickey(2018)Mej{\'{i}}a-Rodr{\'{i}}guez, and Trickey]{SCANL2}
Mej{\'{i}}a-Rodr{\'{i}}guez,~D.; Trickey,~S.~B. {Deorbitalized meta-GGA
  Exchange-Correlation Functionals in Solids}. \emph{Phys. Rev. B}
  \textbf{2018}, \emph{98}, 115161\relax
\mciteBstWouldAddEndPuncttrue
\mciteSetBstMidEndSepPunct{\mcitedefaultmidpunct}
{\mcitedefaultendpunct}{\mcitedefaultseppunct}\relax
\EndOfBibitem
\bibitem[Mej{\'{i}}a-Rodr{\'{i}}guez and
  Trickey(2020)Mej{\'{i}}a-Rodr{\'{i}}guez, and Trickey]{rrSCANL}
Mej{\'{i}}a-Rodr{\'{i}}guez,~D.; Trickey,~S.~B. {Meta-GGA Performance in Solids
  at Almost GGA Cost}. \emph{Phys. Rev. B} \textbf{2020}, \emph{102},
  121109(R)\relax
\mciteBstWouldAddEndPuncttrue
\mciteSetBstMidEndSepPunct{\mcitedefaultmidpunct}
{\mcitedefaultendpunct}{\mcitedefaultseppunct}\relax
\EndOfBibitem
\bibitem[Apr{\`{a}} \latin{et~al.}(2020)Apr{\`{a}}, Bylaska, de~Jong, Govind,
  Kowalski, Straatsma, Valiev, van Dam, Alexeev, Anchell, Anisimov, Aquino,
  Atta-Fynn, Autschbach, Bauman, Becca, Bernholdt, Bhaskaran-Nair, Bogatko,
  Borowski, Boschen, Brabec, Bruner, Cau{\"{e}}t, Chen, Chuev, Cramer, Daily,
  Deegan, Dunning, Dupuis, Dyall, Fann, Fischer, Fonari, Fr{\"{u}}chtl,
  Gagliardi, Garza, Gawande, Ghosh, Glaesemann, G{\"{o}}tz, Hammond, Helms,
  Hermes, Hirao, Hirata, Jacquelin, Jensen, Johnson, J{\'{o}}nsson, Kendall,
  Klemm, Kobayashi, Konkov, Krishnamoorthy, Krishnan, Lin, Lins, Littlefield,
  Logsdail, Lopata, Ma, Marenich, del Campo, Mejia-Rodriguez, Moore, Mullin,
  Nakajima, Nascimento, Nichols, Nichols, Nieplocha, Otero-de-la Roza, Palmer,
  Panyala, Pirojsirikul, Peng, Peverati, Pittner, Pollack, Richard, Sadayappan,
  Schatz, Shelton, Silverstein, Smith, Soares, Song, Swart, Taylor, Thomas,
  Tipparaju, Truhlar, Tsemekhman, {Van Voorhis}, V{\'{a}}zquez-Mayagoitia,
  Verma, Villa, Vishnu, Vogiatzis, Wang, Weare, Williamson, Windus,
  Woli{\'{n}}ski, Wong, Wu, Yang, Yu, Zacharias, Zhang, Zhao, and
  Harrison]{NWChem2020}
Apr{\`{a}},~E.; Bylaska,~E.~J.; de~Jong,~W.~A.; Govind,~N.; Kowalski,~K.;
  Straatsma,~T.~P.; Valiev,~M.; van Dam,~H. J.~J.; Alexeev,~Y.; Anchell,~J.
  \latin{et~al.}  {NWChem: Past, present, and future}. \emph{J. Chem. Phys.}
  \textbf{2020}, \emph{152}, 184102\relax
\mciteBstWouldAddEndPuncttrue
\mciteSetBstMidEndSepPunct{\mcitedefaultmidpunct}
{\mcitedefaultendpunct}{\mcitedefaultseppunct}\relax
\EndOfBibitem
\bibitem[Weigend and Ahlrichs(2005)Weigend, and Ahlrichs]{def2}
Weigend,~F.; Ahlrichs,~R. {Balanced basis sets of split valence, triple zeta
  valence and quadruple zeta valence quality for H to Rn: Design and assessment
  of accuracy}. \emph{Phys. Chem. Chem. Phys.} \textbf{2005}, \emph{7},
  3297--3305\relax
\mciteBstWouldAddEndPuncttrue
\mciteSetBstMidEndSepPunct{\mcitedefaultmidpunct}
{\mcitedefaultendpunct}{\mcitedefaultseppunct}\relax
\EndOfBibitem
\bibitem[Lebedev(1975)]{Lebedev75}
Lebedev,~V. Values of the nodes and weights of quadrature formulas of
  Gauss-Markov type for a sphere from the ninth to seventeenth order of
  accuracy that are invariant with respect to an octahedron group with
  inversion. \emph{Zh. Vychisl. Mat. mat. Fiz.} \textbf{1975}, \emph{15},
  48\relax
\mciteBstWouldAddEndPuncttrue
\mciteSetBstMidEndSepPunct{\mcitedefaultmidpunct}
{\mcitedefaultendpunct}{\mcitedefaultseppunct}\relax
\EndOfBibitem
\bibitem[Lebedev(1976)]{Lebedev76}
Lebedev,~V. Quadratures on the sphere. \emph{Zh. Vychisl. Mat. Mat. Fiz.}
  \textbf{1976}, \emph{16}, 293\relax
\mciteBstWouldAddEndPuncttrue
\mciteSetBstMidEndSepPunct{\mcitedefaultmidpunct}
{\mcitedefaultendpunct}{\mcitedefaultseppunct}\relax
\EndOfBibitem
\bibitem[Lebedev(1977)]{Lebedev77}
Lebedev,~V. Spherical quadrature formulas exact to orders 25-29. \emph{Sibirsk.
  Mat. Zh.} \textbf{1977}, \emph{18}, 132\relax
\mciteBstWouldAddEndPuncttrue
\mciteSetBstMidEndSepPunct{\mcitedefaultmidpunct}
{\mcitedefaultendpunct}{\mcitedefaultseppunct}\relax
\EndOfBibitem
\bibitem[Lebedev and Skorokhodov(1992)Lebedev, and
  Skorokhodov]{LebedevSkorokhodov}
Lebedev,~V.; Skorokhodov,~A. Quadrature formulas for a sphere of orders 41, 47
  and 53. \emph{Dokl. Akad. Nauk} \textbf{1992}, \emph{324}, 519\relax
\mciteBstWouldAddEndPuncttrue
\mciteSetBstMidEndSepPunct{\mcitedefaultmidpunct}
{\mcitedefaultendpunct}{\mcitedefaultseppunct}\relax
\EndOfBibitem
\bibitem[Lebedev(1994)]{Lebedev94}
Lebedev,~V. A quadrature formula for the sphere of 59th algebraic order of
  accuracy. \emph{Dokl. Akad. Nauk} \textbf{1994}, \emph{338}, 454\relax
\mciteBstWouldAddEndPuncttrue
\mciteSetBstMidEndSepPunct{\mcitedefaultmidpunct}
{\mcitedefaultendpunct}{\mcitedefaultseppunct}\relax
\EndOfBibitem
\bibitem[Lebedev and Laikov(1999)Lebedev, and Laikov]{LebedevLaikov}
Lebedev,~V.; Laikov,~D. Quadrature formula for the sphere of 131st algebraic
  order of accuracy. \emph{Dokl. Akad. Nauk} \textbf{1999}, \emph{366},
  741\relax
\mciteBstWouldAddEndPuncttrue
\mciteSetBstMidEndSepPunct{\mcitedefaultmidpunct}
{\mcitedefaultendpunct}{\mcitedefaultseppunct}\relax
\EndOfBibitem
\bibitem[Weigend(2006)]{Weigend2006}
Weigend,~F. {Accurate Coulomb-fitting basis sets for H to Rn}. \emph{Phys.
  Chem. Chem. Phys.} \textbf{2006}, \emph{8}, 1057--1065\relax
\mciteBstWouldAddEndPuncttrue
\mciteSetBstMidEndSepPunct{\mcitedefaultmidpunct}
{\mcitedefaultendpunct}{\mcitedefaultseppunct}\relax
\EndOfBibitem
\bibitem[Whitten(1973)]{Whitten1973}
Whitten,~J. {Coulombic potential energy integrals and approximations}. \emph{J.
  Chem. Phys.} \textbf{1973}, \emph{58}, 4496\relax
\mciteBstWouldAddEndPuncttrue
\mciteSetBstMidEndSepPunct{\mcitedefaultmidpunct}
{\mcitedefaultendpunct}{\mcitedefaultseppunct}\relax
\EndOfBibitem
\bibitem[Dunlap \latin{et~al.}(1979)Dunlap, Conolly, and
  Sabin]{DunlapConollySabin1979}
Dunlap,~B.; Conolly,~J.; Sabin,~J. {On some approximations in applications of
  X$\alpha$ theory}. \emph{J. Chem. Phys.} \textbf{1979}, \emph{71},
  3396--3402\relax
\mciteBstWouldAddEndPuncttrue
\mciteSetBstMidEndSepPunct{\mcitedefaultmidpunct}
{\mcitedefaultendpunct}{\mcitedefaultseppunct}\relax
\EndOfBibitem
\bibitem[Grimme \latin{et~al.}(2010)Grimme, Antony, Ehrlich, and Krieg]{DFTD3}
Grimme,~S.; Antony,~J.; Ehrlich,~S.; Krieg,~H. {A consistent and accurate
  \emph{ab initio} parametrization of density functional dispersion (DFT-D) for
  the 94 elements H-Pu}. \emph{J. Chem. Phys.} \textbf{2010}, \emph{132},
  154104\relax
\mciteBstWouldAddEndPuncttrue
\mciteSetBstMidEndSepPunct{\mcitedefaultmidpunct}
{\mcitedefaultendpunct}{\mcitedefaultseppunct}\relax
\EndOfBibitem
\bibitem[Grimme \latin{et~al.}(2011)Grimme, Ehrlich, and Goerigk]{DFTD3BJ}
Grimme,~S.; Ehrlich,~S.; Goerigk,~L. {Effect of the damping function in
  dispersion corrected density functional theory}. \emph{J. Comput. Chem.}
  \textbf{2011}, \emph{32}, 1456--1465\relax
\mciteBstWouldAddEndPuncttrue
\mciteSetBstMidEndSepPunct{\mcitedefaultmidpunct}
{\mcitedefaultendpunct}{\mcitedefaultseppunct}\relax
\EndOfBibitem
\bibitem[Brandenburg \latin{et~al.}(2016)Brandenburg, Bates, Sun, and
  Perdew]{BrandenburgEtAl16}
Brandenburg,~J.~G.; Bates,~J.~E.; Sun,~J.; Perdew,~J.~P. {Benchmark tests of a
  strongly constrained semilocal functional with a long-range dispersion
  correction}. \emph{Phys. Rev. B.} \textbf{2016}, \emph{94}, 115144\relax
\mciteBstWouldAddEndPuncttrue
\mciteSetBstMidEndSepPunct{\mcitedefaultmidpunct}
{\mcitedefaultendpunct}{\mcitedefaultseppunct}\relax
\EndOfBibitem
\bibitem[Rado\'n(2019)]{Radon2019}
Rado\'n,~M. Benchmarking quantum chemistry methods for spin-state energetics of
  iron complexes against quantitative experimental data. \emph{Phys. Chem.
  Chem. Phys.} \textbf{2019}, \emph{21}, 4854\relax
\mciteBstWouldAddEndPuncttrue
\mciteSetBstMidEndSepPunct{\mcitedefaultmidpunct}
{\mcitedefaultendpunct}{\mcitedefaultseppunct}\relax
\EndOfBibitem
\bibitem[Kresse and Hafner(1993)Kresse, and Hafner]{vasp}
Kresse,~G.; Hafner,~J. {Ab initio molecular dynamics for liquid metals}.
  \emph{Phys. Rev. B} \textbf{1993}, \emph{47}, 558\relax
\mciteBstWouldAddEndPuncttrue
\mciteSetBstMidEndSepPunct{\mcitedefaultmidpunct}
{\mcitedefaultendpunct}{\mcitedefaultseppunct}\relax
\EndOfBibitem
\bibitem[Kresse and Hafner(1994)Kresse, and Hafner]{vasp2}
Kresse,~G.; Hafner,~J. {Ab initio molecular-dynamics simulation of the
  liquid-metal-amorphous-semiconductor transition in germanium}. \emph{Phys.
  Rev. B} \textbf{1994}, \emph{49}, 14251\relax
\mciteBstWouldAddEndPuncttrue
\mciteSetBstMidEndSepPunct{\mcitedefaultmidpunct}
{\mcitedefaultendpunct}{\mcitedefaultseppunct}\relax
\EndOfBibitem
\bibitem[Kresse and Furthm{\"{u}}ller(1996)Kresse, and
  Furthm{\"{u}}ller]{vasp3}
Kresse,~G.; Furthm{\"{u}}ller,~J. {Efficiency iterative schemes for ab initio
  total-energy calculations using a plane-wave basis set}. \emph{Phys. Rev. B}
  \textbf{1996}, \emph{54}, 11169\relax
\mciteBstWouldAddEndPuncttrue
\mciteSetBstMidEndSepPunct{\mcitedefaultmidpunct}
{\mcitedefaultendpunct}{\mcitedefaultseppunct}\relax
\EndOfBibitem
\bibitem[vas(1996)]{vasp4}
Efficiency of ab-initio total energy calculations for metals and semiconductors
  using a plane-wave basis set. \emph{Comput. Mat. Sci.} \textbf{1996},
  \emph{6}, 15\relax
\mciteBstWouldAddEndPuncttrue
\mciteSetBstMidEndSepPunct{\mcitedefaultmidpunct}
{\mcitedefaultendpunct}{\mcitedefaultseppunct}\relax
\EndOfBibitem
\bibitem[Blochl(1994)]{paw1}
Blochl,~P.~E. Projector augmented-wave method. \emph{Phys. Rev. B}
  \textbf{1994}, \emph{50}, 17953\relax
\mciteBstWouldAddEndPuncttrue
\mciteSetBstMidEndSepPunct{\mcitedefaultmidpunct}
{\mcitedefaultendpunct}{\mcitedefaultseppunct}\relax
\EndOfBibitem
\bibitem[Kresse and Joubert(1999)Kresse, and Joubert]{paw2}
Kresse,~G.; Joubert,~D. From ultrasoft pseudopotentials to the projector
  augmented-wave method. \emph{Phys. Rev. B} \textbf{1999}, \emph{59},
  1758\relax
\mciteBstWouldAddEndPuncttrue
\mciteSetBstMidEndSepPunct{\mcitedefaultmidpunct}
{\mcitedefaultendpunct}{\mcitedefaultseppunct}\relax
\EndOfBibitem
\bibitem[Hinz \latin{et~al.}(2020)Hinz, Karasiev, Hu, Zaghoo,
  Mej{\'i}a-Rodr{\'i}guez, Trickey, and Calder{\'i}n]{HinzEtAl2020}
Hinz,~J.; Karasiev,~V.~V.; Hu,~S.~X.; Zaghoo,~M.; Mej{\'i}a-Rodr{\'i}guez,~D.;
  Trickey,~S.~B.; Calder{\'i}n,~L. {Fully consistent density functional theory
  determination of the insulator-metal transition boundary in warm dense
  hydrogen}. \emph{Phys. Rev. Research} \textbf{2020}, \emph{2}, 032065\relax
\mciteBstWouldAddEndPuncttrue
\mciteSetBstMidEndSepPunct{\mcitedefaultmidpunct}
{\mcitedefaultendpunct}{\mcitedefaultseppunct}\relax
\EndOfBibitem
\bibitem[Zhang \latin{et~al.}(2020)Zhang, Zhang, and Singh]{ZhangZhangSingh}
Zhang,~Y.; Zhang,~W.; Singh,~D.~J. {Localization in the SCAN meta-generalized
  gradient approximation functional leading to broken symmetry ground states
  for graphene and benzene}. \emph{Phys. Chem. Chem. Phys.} \textbf{2020},
  \emph{22}, 19585--19591\relax
\mciteBstWouldAddEndPuncttrue
\mciteSetBstMidEndSepPunct{\mcitedefaultmidpunct}
{\mcitedefaultendpunct}{\mcitedefaultseppunct}\relax
\EndOfBibitem
\bibitem[Perdew()]{PerdewPC}
Perdew,~J.~P. Private communication, 2020.\relax
\mciteBstWouldAddEndPunctfalse
\mciteSetBstMidEndSepPunct{\mcitedefaultmidpunct}
{}{\mcitedefaultseppunct}\relax
\EndOfBibitem
\bibitem[Casey and Leopold(1993)Casey, and Leopold]{CaseyLeopold1993}
Casey,~S.~M.; Leopold,~D.~G. Negative ion photoelectron spectroscopy of Cr$_2$.
  \emph{J. Phys. Chem.} \textbf{1993}, \emph{97}, 816--830\relax
\mciteBstWouldAddEndPuncttrue
\mciteSetBstMidEndSepPunct{\mcitedefaultmidpunct}
{\mcitedefaultendpunct}{\mcitedefaultseppunct}\relax
\EndOfBibitem
\bibitem[Simard \latin{et~al.}(1998)Simard, Lebeault-Dorget, Marijnissen, and
  {ter Meulen}]{SimardEtAl1998}
Simard,~B.; Lebeault-Dorget,~M.-A.; Marijnissen,~A.; {ter Meulen},~J.~J.
  Photoionization spectroscopy of dichromium and dimolybdenum: Ionization
  potentials and bond energies. \emph{J. Chem. Phys.} \textbf{1998},
  \emph{108}, 9668\relax
\mciteBstWouldAddEndPuncttrue
\mciteSetBstMidEndSepPunct{\mcitedefaultmidpunct}
{\mcitedefaultendpunct}{\mcitedefaultseppunct}\relax
\EndOfBibitem
\bibitem[Vacoillie \latin{et~al.}(2016)Vacoillie, \r{A}ke Malmqvist, and
  Veryazov]{Vanco2016}
Vacoillie,~S.; \r{A}ke Malmqvist,~P.; Veryazov,~V. Potential energy surface of
  the chromium dimer re-re-visited with multiconfigurational perturbation
  theory. \emph{J. Chem. Theory Comput.} \textbf{2016}, \emph{12},
  1647--1655\relax
\mciteBstWouldAddEndPuncttrue
\mciteSetBstMidEndSepPunct{\mcitedefaultmidpunct}
{\mcitedefaultendpunct}{\mcitedefaultseppunct}\relax
\EndOfBibitem
\end{mcitethebibliography}

\begin{tocentry}
\centering
\vspace{0.1cm}
\includegraphics[width=2.5in]{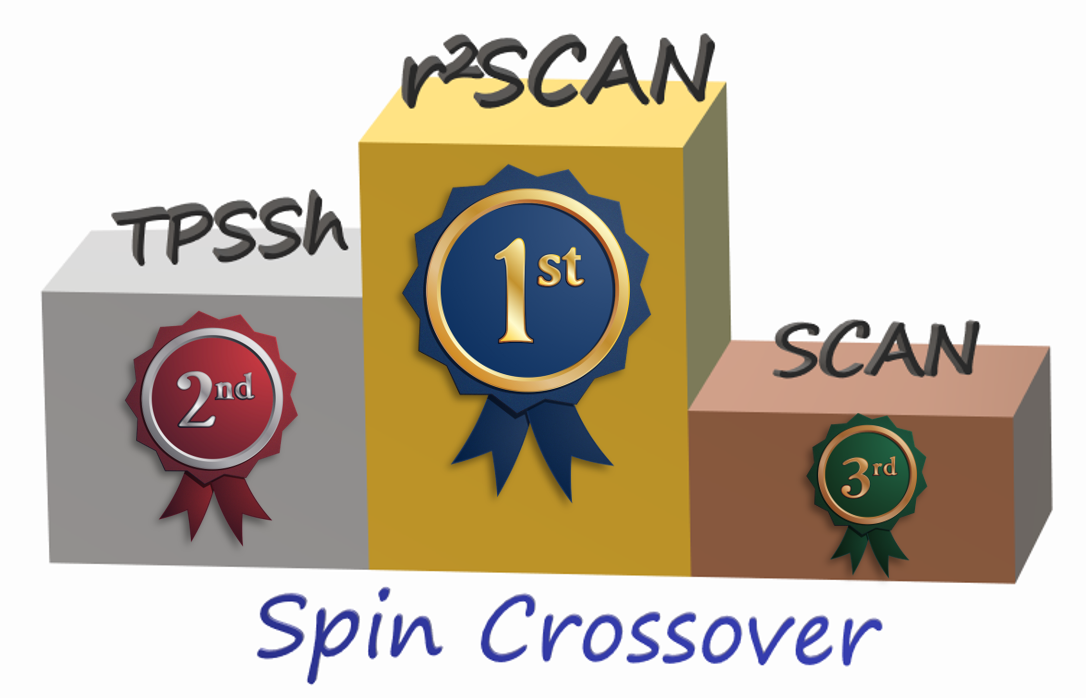}
\end{tocentry}

\end{document}